\documentclass[10pt,letterpaper]{article}
\usepackage[top=0.85in,footskip=0.75in]{geometry}
\usepackage{amsmath,amssymb}
\usepackage{changepage}
\usepackage[utf8x]{inputenc}
\usepackage{textcomp,marvosym}
\usepackage{cite}
\usepackage{nameref}
\usepackage{microtype}
\DisableLigatures[f]{encoding = *, family = * }
\usepackage[table]{xcolor}
\usepackage{array}
\usepackage{graphicx}
\usepackage{setspace} 
\usepackage[aboveskip=1pt,labelfont=bf,labelsep=period,justification=raggedright,singlelinecheck=off]{caption}

\makeatletter
\renewcommand{\@biblabel}[1]{\quad#1.}
\makeatother

\date{}

\begin{document}
\vspace*{0.2in}

\begin{flushleft}
{\Large
\textbf\newline{Reconfiguring motor circuits for a joint manual and BCI task}
}
\newline
\\
Benjamin Lansdell\textsuperscript{1 *+},
Ivana Milovanovic\textsuperscript{2+},
Cooper Mellema\textsuperscript{3},
Eberhard E Fetz\textsuperscript{3,4,5},
Adrienne L Fairhall\textsuperscript{3,4,5},
Chet T Moritz\textsuperscript{6,2,3,4,5},
\\
\bigskip
\textbf{1} Applied Mathematics, University of Washington, Seattle, WA, USA
\\
\textbf{2} Rehabilitation Medicine, University of Washington, Seattle, WA, USA
\\
\textbf{3} Physiology and Biophysics, University of Washington, Seattle, WA, USA
\\
\textbf{4} Center for Neurotechnology, University of Washington, Seattle, WA, USA
\\
\textbf{5} UW Institute for Neural Engineering, University of Washington, Seattle, WA, USA
\\
\textbf{6} Electrical \& Computer Engineering, University of Washington, Seattle, WA, USA
\\
\bigskip
* lansdell@uw.edu\\
+ These authors contributed equally.

\end{flushleft}
\section*{Abstract}
Designing brain-computer interfaces (BCIs) that can be used in conjunction with ongoing motor behavior requires an understanding of how neural activity co-opted for brain control interacts with existing neural circuits. For example, BCIs may be used to regain lost motor function after stroke. This requires that neural activity controlling unaffected limbs is dissociated from activity controlling the BCI. In this study we investigated how primary motor cortex accomplishes simultaneous BCI control and motor control in a task that explicitly required both activities to be driven from the same brain region (i.e. a dual-control task). Single-unit activity was recorded from intracortical, multi-electrode arrays while a non-human primate performed this dual-control task. Compared to activity observed during naturalistic motor control, we found that both units used to drive the BCI directly (control units) and units that did not directly control the BCI (non-control units) significantly changed their tuning to wrist torque. Using a measure of effective connectivity, we observed that control units decrease their connectivity. Through an analysis of variance we found that the intrinsic variability of the control units has a significant effect on task proficiency. When this variance is accounted for, motor cortical activity is flexible enough to perform novel BCI tasks that require active decoupling of natural associations to wrist motion. This study provides insight into the neural activity that enables a dual-control brain-computer interface.

\section*{Introduction}

Broad application of brain-computer interfaces (BCIs) for control of neural prostheses requires understanding how brain circuits can simultaneously engage in competing tasks. In the case of partial paralysis resulting from stroke, restoring function with a BCI-controlled stimulator would require coordination between control of the BCI and activity governing the production of residual movement. Further, allowing direct brain control of external devices by healthy subjects has many potential industrial applications.

Both invasive and non-invasive brain computer interfaces have potential to improve quality of life for people with sensorimotor challenges. Non-invasive BCIs have demonstrated significant success in rehabilitation and communication applications \cite{Waldert2016,Kalagi2017,Bundy2012,Nijboer2008}. However, implanted BCIs provide a high bandwidth for control of continuous decoding of movement intention. Implanted BCIs therefore have the potential to realize the most ambitious applications of BCIs.  

Typical intra-cortical BCIs in humans and animals operate using population decoding based on real or imagined movement \cite{Gilja2015, Aflalo2015, Hochberg2012}. In particular, brain-control mappings that make use of activity observed during the natural motor repertoire have been shown to be most effective \cite{Hwang2013, Sadtler2014}. However, co-opting natural mappings between population activity and movement may pose a challenge when BCI control is required in parallel with ongoing motor activity. We refer to BCIs designed to be used simultaneously with natural motor output as dual-control BCIs. In this article we investigate representation and performance in a dual-control BCI.

One strategy for designing dual-control BCIs is to take advantage of conditioning paradigms that allow for volitional control of neural signals \cite{Fetz1969, Fetz1973, Fetz2007, Moritz2008, Moritz2011, Aflalo2015}. Previous studies show that effective control can be achieved even when selected control units show little prior tuning to motor output, providing a large candidate population of control signals \cite{Fetz1975}. Such neural interfaces may thus be adapted to novel simultaneous-control tasks which require the coordination between networks of neurons responsible for movement and brain control \cite{Milovanovic2015}. However, while subpopulations of neurons encoding ipsilateral motion are a candidate control source for the BCI following stroke \cite{Sharma2009}, a robust interface may require recording from the larger population of neurons that are involved in natural, contralateral control. In such cases, networks engaged for brain control and motor control may overlap. Proficient use of a dual-control BCI thus requires that the networks coordinate or dissociate in a way that supports independent control of both natural movement and the BCI.

Previous studies demonstrate that motor cortical activity is flexible enough to operate a dual-control BCI. In our previous dual-control BCI study \cite{Milovanovic2015}, single units that are wrist flexion/extension-tuned in a motor task could be used to control an independent axis in a task that requires concurrent wrist flexion/extension. In this task, the monkey was able to learn a dual-control task at a rate well above chance. Performance, measured as successful trials per minute, improved throughout a session, indicating that the monkey adapted to the task over a matter of minutes. A related study demonstrates that similar concurrent-use BCIs are possible in human subjects using ECoG signals \cite{Bashford2016}. Another primate study shows robustness to interference from native motor networks \cite{Orsborn2014}. These studies suggest that dissociation of BCI and motor control networks is indeed possible for the purposes of controlling a dual-control BCI.

While our previous study demonstrates that it is possible to dissociate motor control from BCI control, the mechanisms of dissociation in the supporting neural networks has not been investigated. How is activity responsible for BCI control coordinated with the activity responsible for ongoing wrist movement during use of a dual-control BCI? 

\begin{figure}[!t]
\centering
\includegraphics[width=3.2in]{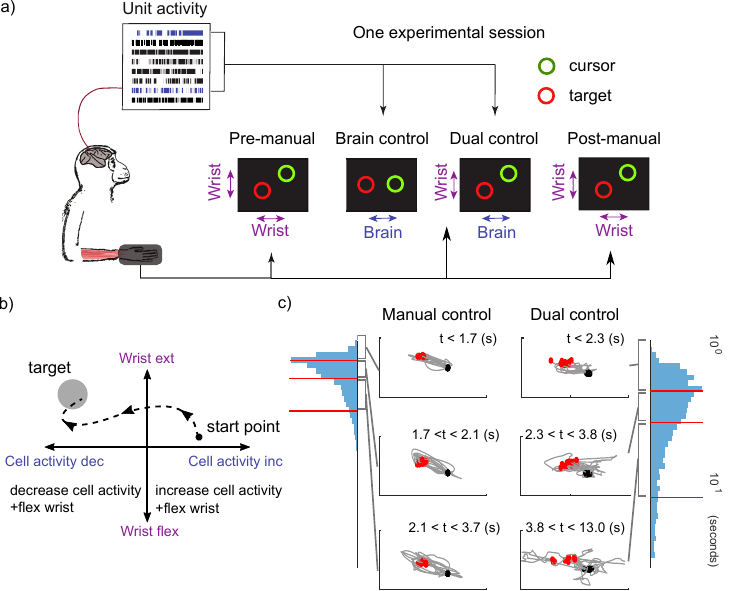}
\caption{Dual-control BCI experimental setup. Analysis is based on data and experiments performed in \cite{Milovanovic2015}.  a) Isolated primary motor cortex unit activity controls a brain-control axis, while contralateral wrist torque determines a manual-control axis. In each experimental session, the monkey first performed a 2D manual wrist task. The monkey then performed a 1D brain-control task, in which cursor velocity was determined by neural activity of two units. The monkey then performed a dual-control task, where one axis is determined by wrist torque and the other by neural activity. Finally, the monkey performed a second 2D manual task. b) Random target pursuit task for dual-control BCI. Targets (grey disk) randomly appear, subject has fixed time to acquire and hold cursor within disk. Thus in the dual-control task, subjects are required to modulate the activity of control units independently of wrist motion. c) Example trial trajectories for manual (left) and dual control (right) tasks. Trajectories are sorted by trial time. All trajectories are shifted and rotated so the start point is on the right (black dot) and the end point is on the left (red dot). Histograms show distribution of trial times (in seconds, on log scale). Red lines highlight the 30\%, 60\% and 90\% percentile trial times, which define the three sub-populations from which sample trials are drawn.}
\label{fig1}
\end{figure}

Here we analyze these questions using the same task and data as our previous study \cite{Milovanovic2015}. Specifically, we address these questions using a dual-control BCI task, where both BCI control and ongoing motor output are driven by the same cortical region (Fig. \ref{fig1}; task and data reported in \cite{Milovanovic2015}). By recording the activity of both control and non-control units, we study how their activity differs between dual-control, manual control, and brain-control tasks. We seek to understand how population activity observed during dual control relates to activity observed during manual- and brain-control tasks.

Using both a linear-tuning analysis and an effective-connectivity analysis, we find that dual-control neural activity is more similar to manual-control neural activity than brain-control neural activity. By design, proficiency at the dual-control task requires units to fire independently of wrist motion. Therefore, we hypothesized that during the task, the activity of control units would dissociate from previously co-tuned units. We find evidence that during dual control, control units dissociate specifically from co-tuned units in a different way than how they dissociate in the brain control task, suggesting that dissociation required by the dual-control task can occur with single-unit specificity. Finally, we searched for factors that predict performance in the dual-control task. Building on previous studies \cite{Hwang2013, Sadtler2014}, we find that measures of single unit intrinsic variability are predictive of dual-control BCI task performance.

\section{Methods}

Our analysis is based on experiments performed in \cite{Milovanovic2015}. We review these methods here. Briefly, one male Macaque nemestrina monkey was trained to perform a random target-pursuit motor task. The monkey was implanted with two 96-channel multi-electrode arrays bilaterally in primary motor cortex. In this study only data from the left hemisphere were used. All procedures were in accordance with National Institutes of Health ‘Guide for the Care and Use of Laboratory Animals’. Further details of surgery and electrode implantation, and behavioral training are detailed in \cite{Milovanovic2015}.

The monkey began each daily session by controlling the cursor with isometric wrist torque in two dimensions (manual control, MC), then progressed to using the aggregate neural activity of two single units to control a cursor in one dimension (brain-control, BC). Subsequently, he used the same neural activity to control the BCI in one dimension, while simultaneously using isometric wrist torque of the contralateral forelimb to control the cursor in a second orthogonal dimension (dual-control, DC; Fig \ref{fig1}).

The random target pursuit task involved the monkey moving the cursor to the target and maintaining the cursor within the target for at least 1s to receive a reward. Targets appeared randomly, and a 0.5s break was provided between trials. For each trial there was a time-out period of 40s. Besides the dimensionality, the task for the manual-control, brain-control and dual-control settings was the same. 

Each day a new pair of units was chosen to control the BCI. If directionally tuned units were used that day, units with approximately opposite preferred directions were selected and paired to control the BCI. Each unit contributed to positive cursor velocity in the direction most closely aligned with its preferred direction. Thus, the contribution of one of the tuned units’ activity was subtracted from the other to determine the cursor velocity. 

In order to ensure decoupling of unit activity and hand control in the dual control task, visual feedback of one of these modalities had to be rotated to achieve independent degrees of freedom on the monitor. Since the monkey was overtrained on the manual control task, we made a deliberate decision to preserve the relation between wrist torque and cursor movement. Instead, we chose to rotate the visual feedback of the units’ firing rate by 90 degrees relative to their preferred direction, even for units with little to no directional preference. In other words, the brain control axis was orthogonal to the preferred direction of the control units. This choice had minimal effect on the changes in tuning angles observed between conditions (Appendix B). The same BCI axis was used for both the brain and dual control trials in a given session. 

We verified performance in dual control was above chance by generating `null' sessions. The trajectories the monkey generated were run through the same task logic, but with a new set of randomly chosen targets unrelated to the targets the monkey actually was presented with. Performance (success rate) on these `null' trials gives a baseline measure of performance expected through random cursor motion alone. We found that 72\% of the sessions had performance significantly above random performance ($\alpha = 0.05$), and most were well above the chance rate. Thus in most sessions the monkey does indeed learn the dual control task. We also see a statistically significant increasing trend in performance throughout the pooled dual control sessions, which is further evidence the animal is learning to perform the task successfully. 

All units with a mean firing rate above 5 Hz recorded from the hemisphere contralateral to arm movement were analyzed. The torque signals were smoothed with a Gaussian kernel of width 300ms and a linear tuning model was fit to each unit’s firing rate as a function of wrist torque. Overall performance was quantified as the number of targets per minute acquired throughout the entire recording.

\subsection{BCI decoder}

The cursor position along the BCI axis, $x_t$, was determined as a linear function of $N$ control neurons’ smoothed activity, $y_t^n$, $N=2$, with gain $\alpha_n$ and a running estimate of baseline firing $b_t^n$:
\begin{equation}
\label{eq:bci}
x_t = \sum_{n=1}^N \alpha_n (y_t^n - b_t^n)
\end{equation}
where the baseline firing rate is defined as
\begin{equation}
b_t^n = b_{t-1}^n(1-\gamma) + \gamma y_t^n
\end{equation}
The baseline update rate is given by $\gamma = 0.001$. Spikes were binned at 60Hz. Spike trains were smoothed with a Gaussian kernel of width 100ms. Here alpha was set to equal magnitude and opposite sign for the two units.

\subsection{Linear tuning model}

The kinematic encoding model to isometric wrist torque is defined for each unit $y_i$ simply through the linear relation
\begin{equation}
y_t^i = \alpha^1 x_{t+\tau}^1 + \alpha^2 x_{t+\tau}^2 + c + \epsilon
\end{equation}
where $x^1$ and $x^2$ are the change in isometric wrist torque along each task axes, for predetermined time lag $\tau$, and for Gaussian noise $\epsilon$. Here an offset of $\tau$ = 75ms was used. Spikes are binned at 25 Hz. This time is consistent with previous studies \cite{Shoham2005} and was, on average, the time lag of the maximum cross-correlation between cursor and neural activity. For most units, a model based on the change in wrist torque was found to yield a higher cross-correlation between wrist motion and neural activity than the wrist torque model, so the results presented here are based the change in wrist torque. Change in wrist torque is computed from torque data using a cubic spline \cite{Shoham2005}, which produces smoother trajectories. We experimented with other encoding models, including models with cursor position and target location. Ultimately these additional factors were not found to significantly add to the quality of the model fits. 

The model is equivalent to a cosine encoding model through a simple transformation:
\begin{equation}
y_t^i=\alpha r \cos(\theta - \theta_{pref}) + c + \epsilon
\end{equation}
for $\alpha = \sqrt{(\alpha^1)^2 + (\alpha^2)^2}$,$\theta_{pref}=\tan^{-1}(\alpha^2 / \alpha^1)$, $r = \sqrt{(x^1)^2 + (x^2)^2}$ and $\theta = \tan^{-1}(x^2/x^1)$. From this we can interpret $\alpha/c$ as a measure of modulation depth. We will refer to $\theta_{pref}$ as tuning angle and $\alpha$ as tuning strength.

To select units for linear tuning analysis, we examined the distribution of $R^2$ values in the manual control task. This showed a discontinuity near $R^2$=0.01. Accordingly, we selected units whose $R^2$ was above 0.01 in the manual control recording. The majority of units chosen according to this heuristic were significantly tuned (according to a t-test on linear coefficients, p < 0.05). This resulted in 411 units being analyzed, 57 of which were control units. The average duration of the manual control sessions used was 6.20 minutes, comprising an average of 124 trials. The average duration of the brain control sessions used was 10.1 minutes, comprising an average of 60.2 trials. The average duration of the dual-control sessions used was 10.3 minutes, comprising an average of 54.5 trials.

\subsection{Assessing effective connectivity with transfer entropy}
 
Transfer entropy \cite{Schreiber2000} is defined as the conditional mutual information between an observed time series $Y$ and the history of a candidate related series $X$, conditional on the history of $Y$
\begin{align}
\notag I_{X\to Y} =& H(Y_t|Y_{t-1},\dots,Y_{t-T}) \\
            &- H(Y_t|Y_{t-1},\dots,Y_{t-T},X_{t-1},\dots,X_{t-T})
\end{align}

As with many information theoretic quantities, $I_{X\to Y}$ is expensive to compute. An approximation tailored for spiking data is used, in particular the transfer entropy toolbox \cite{Ito2011}.

The method was validated using a synthetically generated dataset and compared with other common effective or functional connectivity measures. Specifically it was compared to linear Granger causality, Poisson-process Granger causality, and correlation. On the synthetic data transfer entropy was most accurate.

Transfer entropy is computed with spikes binned at 5ms intervals. Six minutes of recording data in each condition is used. Connections extend up to 30 time-bins into the past (150ms).

The connectivity analysis is performed as follows. Within each session, between recording conditions (manual to brain control and manual to dual control), changes in connectivity are compared for different populations of pairs of units. Recalling that transfer entropy is a directed measure of connectivity, the populations of unit pairs are: to control units and to non-control units; co-tuned units to control units and non-co-tuned units to control units; and co-tuned units to non-control units and non-co-tuned units to non-control units.

For a given session, generally no more than two pairs of units could be identified within 45 degrees of the unit of interest (e.g. the control unit). Thus in comparisons involving co-tuned units, for each session only the top two co-tuned units were selected. These are defined as the two unit pairs having the closest tuning angle. All other pairs of units in the session were classified as not co-tuned (defined this way, most cotuned units are within 45 degrees of each other, while most non-cotuned units are more than 45 degrees apart, Fig 4a). In non-co-tuned populations generally many more than two pairs of units can be identified within a session. Thus to ensure these larger populations are better sampled, and balanced with populations of co-tuned units, the selection of two units is bootstrap sampled 50 times for each session. Such generated populations are combined over all sessions. All co-tuned populations are selected on the basis of tuning during manual control recordings. This resulted in approximately 4400 connections being analyzed over the 99 sessions.

\subsection{Granger-causality cursor control metric}

Given the form of the linear decoding model (Eq 1), it is reasonable to model the cursor trajectory as a moving average time series:
$$x_t=\sum_{n=1}^N\sum_{p=1}^Pk_py_{t-p}^n+c$$
for some model order $P$ to be determined. Spikes are binned at 25Hz, and cursor data is smoothed with a cubic spline, as per the linear tuning model \cite{Shoham2005}.

From the linear model we compute the maximum likelihood estimate given both neurons, $\hat{\beta}_{MLE}^i$, and the maximum likelihood estimate  with neuron $j$ withheld, $\hat{\beta}_{MLE}^{i\setminus j}$. The difference between the two log-likelihoods of the MLE models provides a Granger-causality type metric \cite{Barnett2014}
$$\mathcal{G}_{j\to i}=2\mathcal{L}(\hat{\beta}_{MLE}^{i};\mathbf{x},\mathbf{y}) - 2\mathcal{L}(\hat{\beta}_{MLE}^{i\setminus j};\mathbf{x},\mathbf{y})$$
that quantifies the influence that one neuron has on the cursor trajectory. The likelihood ratio is often used as the basis for hypothesis testing, though the quantity itself can be interpreted as a form of transfer entropy, provided the data is Gaussian distributed \cite{Barnett2009}.

The Granger causality toolbox MVGC \cite{Barnett2014} was used to perform this analysis. The approach is well suited since the cursor position is, by definition, a moving average model of the neural data. The method assumes a covariance stationary time series. By construction (Eq 1) the cursor trajectory has this property, provided that the neural data is second-order stationary. This was confirmed by checking the spectral radius of the vector autoregressive model was less than one.

To verify that $\mathcal{G}_{j\to i}$ has quantitative meaning we generated synthetic data from neural recordings. A cursor trajectory was generated as the weighted sum of the two units’ decoded trajectories, according to (Eq 1). Following this we computed $\mathcal{G}_{j\to i}$ for different weightings. A strong monotonic relationship exists between $\mathcal{G}_{j\to i}$ and the weight that determines how much each unit contributed to the cursor trajectories (data not shown), demonstrating that the control metric can recover which unit contributes more to cursor control.

For the variability analysis, 198 control units are chosen from the 99 sessions.

\section{Results}

\subsection{Single unit analysis}

Given that the dual-control task requires both brain control and manual control, we first sought to understand how the population activity under dual control relates to that observed during the manual-control and brain-control tasks. To do this, a monkey performed the following trials. Each session began with a block of manual-control trials, then brain-control, then dual-control and concluded with a second block of manual-control trials. During each session a fixed population of neurons was identified, allowing changes in their tuning and effective connectivity properties between tasks to be calculated. We selected neurons that were active ($>5$Hz) in all trial blocks. We identified 411 such units throughout the 99 sessions, of which 57 were control units.

\begin{figure}[!t]
\centering
\includegraphics[width=3in]{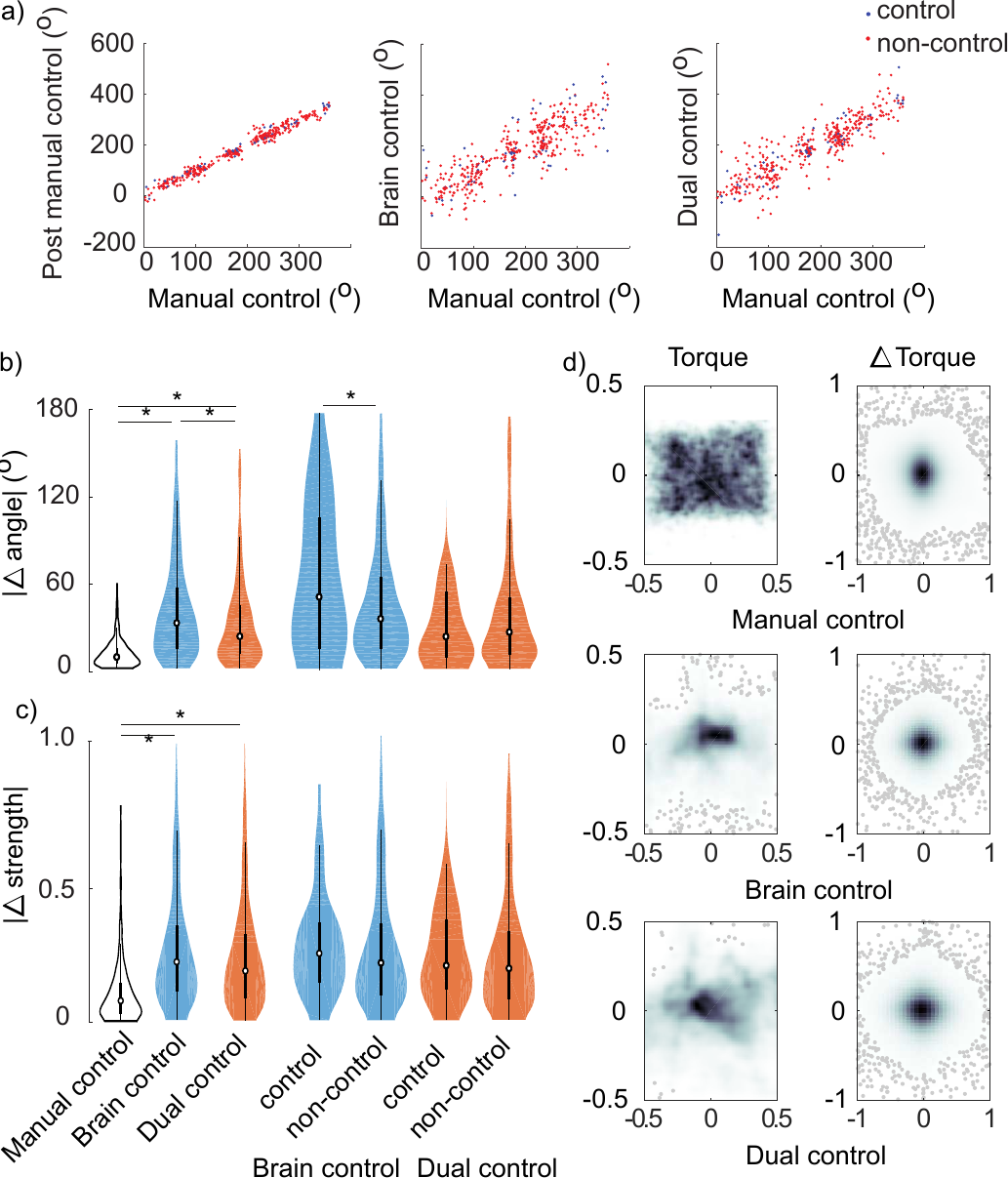}
\caption{a) Tuning angle for control and non-control units between tasks. b) Mean absolute differences in tuning angle between manual control task and both brain control and dual control tasks are significant. (two sided t-test; all $p \ll 0.001$). Mean change in angle between control and non-control units is significantly different  in brain-control (two sided t-test: $p = 0.008$) but not dual-control task. Violin plots show density estimate, along with median and interquartile range. c) Mean absolute differences in tuning strength are significant between manual control task and both brain control and dual control. (two sided t-test: $p\ll0.001$) d) Density plot of both torque and change in torque data for wrist movements in manual-, brain- and dual-control tasks. Outliers (grey dots) represent 5\% of data.}
\label{fig2}
\end{figure}

We fit a linear tuning model for each neuron to wrist movement and an effective connectivity model between recorded units, estimated using transfer entropy \cite{Schreiber2000}. Movement was unrestrained in both dual-control and brain-control trials, allowing each neuron’s association to wrist motion to be estimated in these conditions. The linear model is based on the change in wrist torque, which is distributed equally in the manual-, brain- and dual-control tasks (Fig \ref{fig2}d). Further, the frequency content of the wrist torque under each condition is similar (Appendix A, Figure 6). This means that differences in tuning between tasks can be attributed to differences in the neural activity, and not simply to differences in the underlying movement distribution for each condition. From the linear model we can estimate a measure of a unit’s preferred tuning angle (termed ‘tuning angle’), and a measure of tuning strength (sometimes called modulation depth; refer to Methods). It is also important to note that transfer entropy characterizes a directed measure of covariation, and does not by itself represent any measure of causal or anatomical connectivity. Nonetheless, effective connectivity is one representation of the joint distribution of observed neural network patterns; differences in effectivity connectivity between conditions does capture a difference in the joint distribution between conditions, which is meaningful. We used these two analyses to investigate the network state when the dual-control task was being performed.

Since the dual-control task involves a manual-control component, we sought to understand how many neurons change their tuning from that observed during manual control. We examine differences between the control and non-control units. In order to do so fairly, we checked whether the distribution of tuning properties were significantly different between control and non-control units, as such differences could bias estimates of control unit specific effects. We found no significant difference in the firing rate or tuning strength of control and non-control units in the manual control recordings (two sided t-test). 

During dual control, tuning angles are more similar to manual control than during brain control (Fig \ref{fig2}a,b). Specifically, we found that during the brain control task, only 56 $\pm$ 0.83\% (mean $\pm$ standard error) of control units and 72 $\pm$ 0.64\% of non-control units retain their tuning angle compared to the manual control task. In dual control, however, 78$\pm$0.67\%  of control units and 79 $\pm$ 0.50\% of the non-control units retained a similar tuning angle as in manual control.  These distributions, computed via linear model with resampling, differ between brain and dual control conditions ($p \ll 0.05$; Wilcoxon ranksum test). Thus, the dual-control task produces activity closer to that observed in manual control than is observed during brain control.

\begin{figure}[!t]
\centering
\includegraphics[width=3in]{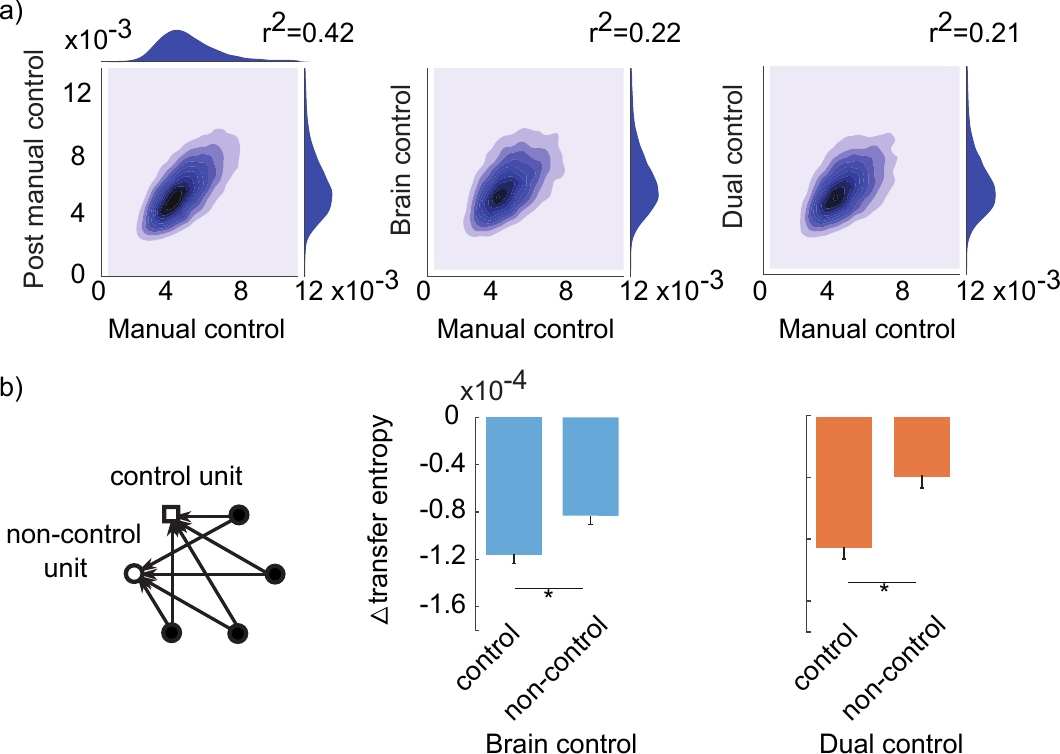}
\caption{a) Changes in effective connectivity between tasks. Transfer entropy between the same pair of units within a session is computed for each task and plotted against one another. Data are pooled over all available sessions. 93\% of units do not significantly change their effective connectivity between pre- and post-manual control recordings (null distribution computed through resampling, $\alpha = 0.05$). b) For each pair of units, summary of effective connectivity changes between manual control and brain-control (left) and manual control and dual-control (right) tasks. Comparisons are made for connections to a control unit (unfilled square) versus to a non-control unit (unfilled circle) from all other recorded units (filled circles). Changes are significantly larger for connections to control units than non-control units (Wilcoxon rank sum; $p \ll 0.001$ ). A negative value means lower connectivity in the brain-control or dual-control condition compared to manual control.}
\label{fig3}
\end{figure}

Further, in the brain-control task there are control-unit specific changes in tuning angle, but not tuning strength. This suggests that some form of dissociation of control unit activity is induced by the BCI (Fig \ref{fig2}b,c). This dissociation is supported by the effective connectivity analysis. While overall there are large changes in effective connectivity between tasks (Fig \ref{fig3}a), control units significantly changed their connectivity more than non-control units in both brain-control and dual-control tasks (Fig \ref{fig3}b). This suggests that both tasks induce some form of dissociation specific to control units, as has been observed in previous single-unit BCI studies \cite{Fetz1973}.

\subsection{Cotuned unit analysis}

Next we focused on changes among populations of co-tuned units during the dual-control task. We define co-tuned units as pairs of units that share preferred direction to wrist motion. Efficiently completing the dual-control task requires the networks associated with wrist control to operate independently from networks associated with BCI control. Thus, for a control unit that is tuned to wrist flexion in a manual control task, any trial that requires wrist flexion and a decrease in control unit activity would decorrelate the control unit’s activity from other units that are also tuned to wrist flexion. The dual-control task does not require that cotuned non-control units  decorrelate from one another. Nor should the dual-control task encourage the control unit to decorrelate from non co-tuned units, since those other units are not activated during wrist flexion. On the basis of these considerations, we hypothesized that in the dual-control task we would observe specifically the control units dissociating from previously co-tuned units from the manual-control task. Further, we predict that this control-unit-specific dissociation would be observed in the dual-control task but not the brain-control task.

\begin{figure*}[!t]
\centering
\includegraphics[width=5.5in]{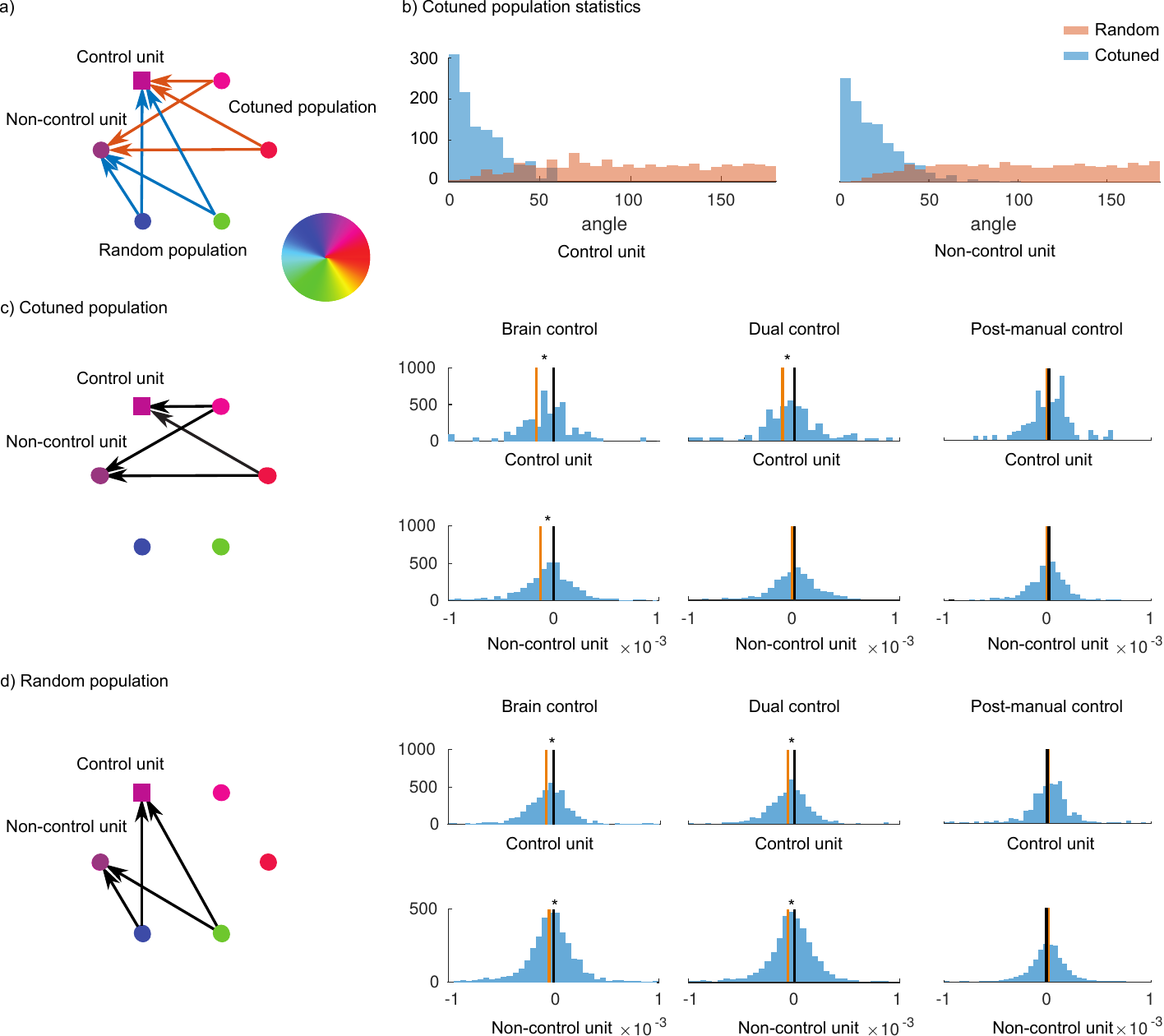}
\caption{Effective connectivity analysis of pairs of cotuned units. a) Effective connectivity is computed for pairs of units for all tasks in a session and then data are pooled over all sessions. Transfer entropy is compared between manual control and brain- or dual-control tasks. Pairs are classified as either co-tuned (indicated by same color) or randomly selected (different colors). Recorded units are compared to either a control or non-control unit. Color wheel indicates tuning angle for example co-tuned and randomly selected units. b) Distribution of tuning angle amongst cotuned population and a non-cotuned population, as identified in the first manual-control task. c) Change in transfer entropy for pairs of units between manual-control and brain-, dual-control or post-manual control tasks, for pairs of units that are cotuned in the manual-control task. The Orange line indicates the mean change from zero (black line; Wilcoxon signed-rank test, * = $p\ll0.05$). A negative value means lower connectivity in the brain-control or dual-control condition. d) Change in transfer entropy for pairs of units between manual-control and brain-, dual- or post-manual control tasks for pairs of units that are not cotuned in the manual-control task.}
\label{fig4}
\end{figure*}

We tested these hypotheses using effective connectivity measured through transfer entropy. Specifically, we asked whether effective connectivity between co-tuned units decreased during the dual-control task for control units more than for non-control units. To answer this question, we computed differences in effective connectivity in the network between manual-control and dual-control trials, and between manual-control and brain-control trials (Fig \ref{fig4}a,b).

We observed similar changes in connectivity between the manual- and brain-control task -- cotuned units dissociated equally, whether a control unit was part of the pair or not. In contrast, between manual- and dual-control trials, only units cotuned with control units dissociated. In fact, connectivity between other cotuned units in the two conditions did not significantly change, suggesting that the dual-control task induces control-unit specific changes to connectivity among cotuned units, supporting our hypothesis (Fig \ref{fig4}c). Further, we find these results are specific to cotuned units -- changes in connectivity between randomly selected units shows no control-unit specific dissociation in either dual- or brain-control tasks (Fig \ref{fig4}d). The changes only occur between the manual control task and either brain or dual control task. Comparing changes in effective connectivity between manual and post-manual control sessions reveals no difference. These results suggest that unit-specific changes occur during the dual-control task in order to support independent control of the BCI.

\subsection{Variability analysis}

Given these insights into the neural activity supporting control of a dual-control BCI, we next sought to understand which features of the recorded primary motor activity, if any, may relate to how well the dual-control task is performed. In our previous study we did not identify differences in tuning that were predictive of performance \cite{Milovanovic2015} -- using both tuned or untuned control units lead to similar performance. We searched for factors within the linear tuning model or the effective connectivity model that may be predictive of task performance or behavior. First, we examined if the 90 degree rotation in BCI axis caused a consistent rotation in the preferred direction of the control units, or any of the recorded units. We found no trend in the change in the tuning of the control units. Second, we also examined if the size of the change in tuning was related to task performance in any way, and also found no relationship. Finally, we examined if the change in connectivity between control units and others was related to task performance. The thinking being that lower connectivity may be related to ‘dissociation’ of the control units from motor control, and that could enable better-dual control performance. Again we found no such relationship. 

However, related BCI studies do show that decoders that utilize the natural motor repertoire are most effective (e.g. \cite{Hwang2013, Sadtler2014}). Thus we hypothesized that similar factors may affect performance in a dual-control BCI.

\begin{figure*}[!t]
\centering
\includegraphics[width=6in]{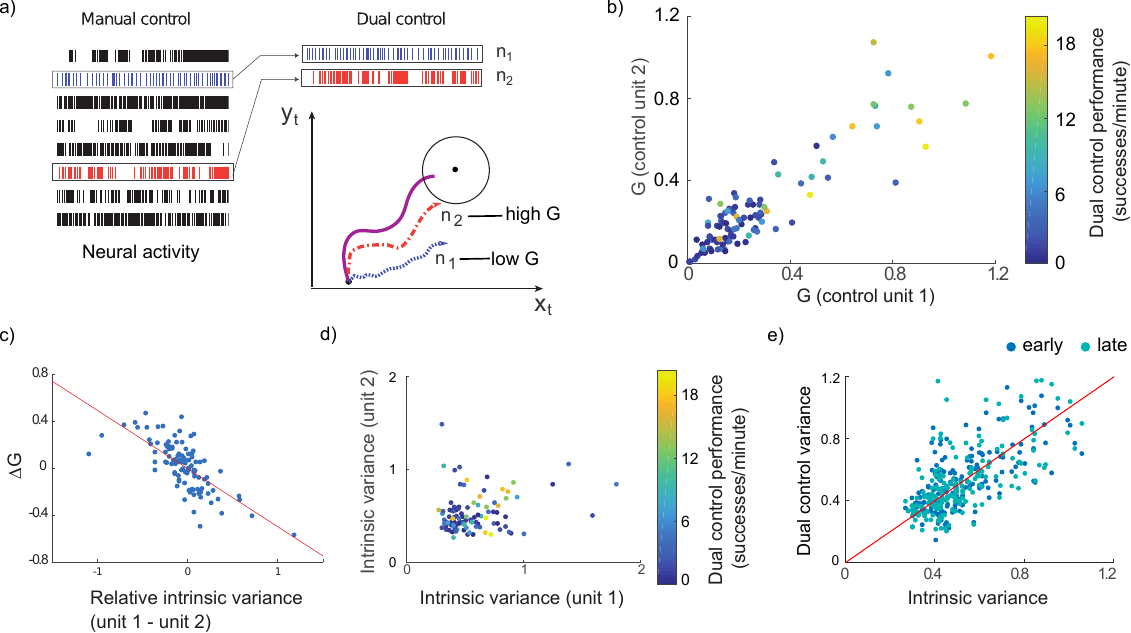}
\caption{a) Control units chosen based on their tuning and activity in manual control recording (left panel) exert different levels of control over cursor in dual control recording (right panel). Dashed lines represent cursor motion contribution from individual control units, while solid purple line represents their aggregate. How much each contributes to cursor motion is measured with a Granger-causality metric, G, which reflects how much the cursor can be predicted on the basis of the activity of each control unit. b) High levels of G are positively correlated with high performance indicated by warmer colors clustering to the right (F-test, $p \ll 0.001$).c) The relative variance of the control units in manual control -- intrinsic variability -- significantly influences the amount of control that each unit exerts over the cursor trajectory in dual control, as identified by the Granger causality analysis ($p \ll 0.001$, $R^2 = 0.500$). d) Intrinsic variability is significantly positively correlated with performance in the dual control task. (F-test, $p = 0.016$). e) Intrinsic variance predicts dual control variance. No significant change in dual control variance occurred between the early and late dual control trials (p = 0.2762, t-test). }
\label{fig5}
\end{figure*}

As our BCI is based on the activity of pairs of units, we performed an analysis of the variability of only these control units. We term the variance of the control unit activity during the manual-control task the intrinsic variability, by analogy with the intrinsic manifold computed in larger neural populations \cite{Sadtler2014}. Specifically, the intrinsic variability is defined as the sample variance estimated from binned spike counts throughout the manual-control task recording. We sought to know if the intrinsic variability of the control units is predictive of how those units will be used for cursor control in the dual-control task. To quantify how much a unit is used to move the cursor, we used a Granger-causality based metric, $\mathcal{G}$, that measures how much the activity of each control unit can predict the cursor trajectory. Due to the design of the decoder, a high influence on the cursor position (high $\mathcal{G}$) is likely related to factors such as the consistency of the firing rate of the unit. That is, the baseline adaptation on the decoder \eqref{eq:bci} effectively decreases the gain on units for which a consistent firing rate can be estimated. We observed that a high $\mathcal{G}$ in each unit is related to high performance (Fig \ref{fig5}b). It is not obvious a priori that units with inconsistent firing rates will be better control units for the dual control task. Further, the unit that contributes more to cursor movement is the unit with the higher intrinsic variability (Fig \ref{fig5}c). Consistent with these results, we observed that high performance in the dual-control task only occurs when the intrinsic variability of the control units is high (Fig \ref{fig5}d). Another factor, control unit firing rate, was not found to significantly relate to dual control performance (data not shown). To investigate if variance increased throughout a block of dual control trials, as might lead to greater control and performance, we split the set of trials into early and late halves. No significant difference between early and late trials was observed (Fig \ref{fig5}e). Variability is thus a strong factor that predicts performance and informs how the dual-control task is performed.

\section{Discussion}

In this study we investigate how neural activity supports a dual-control BCI, where natural movement controls one dimension and brain activity controls a second dimension.

\subsection{Differences between linear tuning in brain-control and dual-control tasks}

We observed that overall tuning to wrist torque are significantly different between manual-control and dual-control. This finding is similar to observations made in a previous primate dual-control study \cite{Orsborn2014}, in which the authors found that producing static force with the arm during the BCI task perturbs the neural map formed during the standard BCI task. Compared to brain-control, we found that tuning angles during dual control were more similar to that observed during natural motion. This suggests that the dual-control task engages native motor networks more than the brain-control task does. It is also possible that natural movement networks are constrained to retain their natural tuning during the dual control task.

In the brain-control task, we observed control-unit specific changes in tuning angle compared to the manual-control task. This effect, however, was absent in the dual-control task. Greater modification of control units tuning angle in the brain-control task appears to be consistent with results from other studies in which visuomotor rotation was applied to a subset of units controlling the cursor \cite{Jarosiewicz2008, Chase2012}. Conversely, between the dual-control and manual-control task, tuning angles changed similarly for both control and non-control units, suggesting that the dual-control task requires population-wide changes in its relation to wrist motion.

Previous BCI studies show differential tuning strength of control versus non-control units \cite{Ganguly2011, Law2014}, while here we report that both control and non-control units undergo similar changes in tuning strength in the brain control and dual control task. Interestingly, Ganguly et al. observed no differential tuning strength of control and non-control groups of units during the initial learning process, but a difference appeared  after 2-3 days of performance with the same units. Similarly, Law et al \cite{Law2014} saw a stronger effect in late compared to early trials. This suggests that the short amount of practice time in the present study for each pair of control units may be insufficient to affect modulation depth.

\subsection{Insights from effective connectivity}

We sought a more detailed characterization of the changes between cotuned units underlying the dual-control task compared with the brain control task. We observed a selective decrease in effective connectivity between co-tuned units and the control unit in dual control, and a non-selective decrease in connectivity between co-tuned units in brain control. This suggests that the dissociation required in the dual control task does affect interactions between individual units in the network. Our results show that co-tuned networks do not change their relation with one another unless needed, when one unit is chosen as a control unit. This is similar to the study of Hwang et al 2013 \cite{Hwang2013}, which shows that neurons maintain their relation to each other when a task is perturbed. Changes in effective connectivity to support motor control tasks have been observed in other studies \cite{Davidson2007}. Thus effective connectivity provides insight into changes that occur while performing a dual-control BCI task.

Effective connectivity, sometimes referred to as functional connectivity, has been used to study properties of motor cortical networks using a range of statistical measures, including directed information \cite{Quinn2011} and Granger causality \cite{Kim2011}. Effective connectivity has provided insight into the operation of both invasive and non-invasive BCIs \cite{Hou2016,Athanasiou2017,Hamedi2016}. In this study we chose to use transfer entropy as a non-parametric measure of effective connectivity, as it performed best on a synthetic validation dataset (data not shown). A drawback of all these methods is that they are unable to distinguish between direct interactions and the effect of hidden, or latent inputs to recorded units. Given the 400 µm electrode separation of the Blackrock Utah array used here, direct synaptic connections between units recorded on different electrodes are unlikely \cite{Smith2009} and all relations between units may be mediated through unobserved connections. A more appropriate model then may be one that separates latent and direct connectivity components \cite{Pfau2013, Semedo2014}.

Given neither the linear tuning analysis or the connectivity analysis reveal factors related to task performance or behavior, the function of the control-unit specific changes in tuning and connective are unclear. Such changes indicate that the recorded population does have a differing role in controlling each task. Specifically what that role is, however, is difficult to say. This is at least in part because neither the linear encoding model or transfer entropy capture causal relationships between either motor output or between neurons. This shortcoming is by no means unique to our study. The only factor we found that did meaningfully relate to task performance was the intrinsic variability, which is consistent with other reports  \cite{Sadtler2014}.

\subsection{Variability constraints}

Finally, in the dual-control task we found that performance is related to each unit’s variability. We found that units with low intrinsic variability do not make a good choice for BCI control units. This is not a surprising result, as the BCI decoder will weigh units with higher variance more heavily. Similar to previous studies \cite{Sadtler2014}, the units with low variability in the manual control task do not increase their variability in the dual control task, even if our results suggest that this would benefit performance.

Previous studies have reported that factors relating to variability are of primary importance when selecting BCI control units. For instance, Yu et al 2009 showed that BCI performance was better when low-dimensional latent dynamics were inferred and utilized, independent of any kind of external movement or task parameter \cite{Yu2009}. A related study showed that trial-by-trial spike predictions based only on the firing rates of simultaneously recorded motor cortex neurons tend to outperform predictions based on external parameters \cite{Stevenson2011}.

Constraints imposed by how well the BCI mapping aligns with the population’s ‘intrinsic manifold’ have also been identified \cite{Sadtler2014, Hwang2013}. In this study the ‘control space’ is only two dimensional, whereas previous studies have explored the concept in larger control spaces consisting of 10s to 100s of units. The low-dimensionality of the present study may facilitate insight into these larger control spaces. In Sadtler et al 2014 \cite{Sadtler2014}, for instance, within-manifold perturbations were less detrimental to performance than ‘outside manifold’ perturbations. In the present study the ‘control space’ is deliberately limited to two dimensions to facilitate insight into how a focused population of neurons must change their activity to accomplish the dual control task. It is possible that the role of variability of control units also explains these previous results. That is, it is possible that the intrinsic manifolds are predominantly defined by axes parallel to highly varying/firing units, and it is these units that dominate the manifold, not linear combinations of units. In other words, correlations between units may not be significant; within/outside manifold perturbations become simply the inclusion or otherwise of varying units that are able to contribute to the control of the BCI.

Along these lines, Athalye et al 2017 \cite{Athalye2017a} focus on individual (‘private’) and shared variability. They report that private variability decreases throughout learning while shared variability increases. Our sessions are likely too short to observe this transition. However the relevance of individual units’ variability in dual-control performance, rather than their coordination, is consistent with these findings. But these ideas warrant further study.

\subsection{Outlook}

Our results demonstrate that dissociation of motor cortical units required by a dual-control task can occur with individual neuron specificity, through task-related changes in both the control and non-control populations. This and previous work \cite{Sadtler2014, Hwang2013} suggests that, provided specific constraints are taken into account, the motor cortex can flexibly adapt to a variety of challenging control tasks. Tasks that require dissociations from established correlations between neural activity and movement can be learned even over the course of a short recording session. Alongside their clinical value as a potential treatment for stroke or other brain injury, BCI paradigms provide insight into the physiological principles that guide motor control \cite{Golub2016} by allowing the relevant populations for motor control and learning to be directly observed. As learning to coordinate separate tasks is an important and general motor control skill, we believe the dual-control BCI promises to yield further insight into the function of primary motor cortex.

\section*{Supporting information}

\appendix

\section{Spectral density of wrist torque for each task}

\begin{figure}[!t]
\centering
\includegraphics[width=2in]{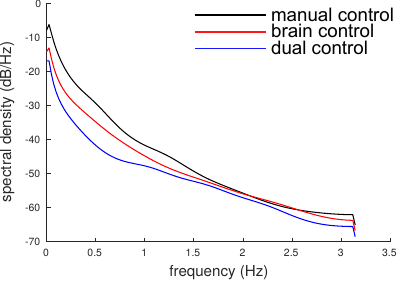}
\caption{Power spectral density of wrist torque over 5 recording sessions under manual, brain and dual control conditions. }
\label{figS1}
\end{figure}

Refer to Fig. \ref{figS1}.

\section{Unrotated sessions tuning analysis}

\begin{figure}[!t]
\centering
\includegraphics[width=3in]{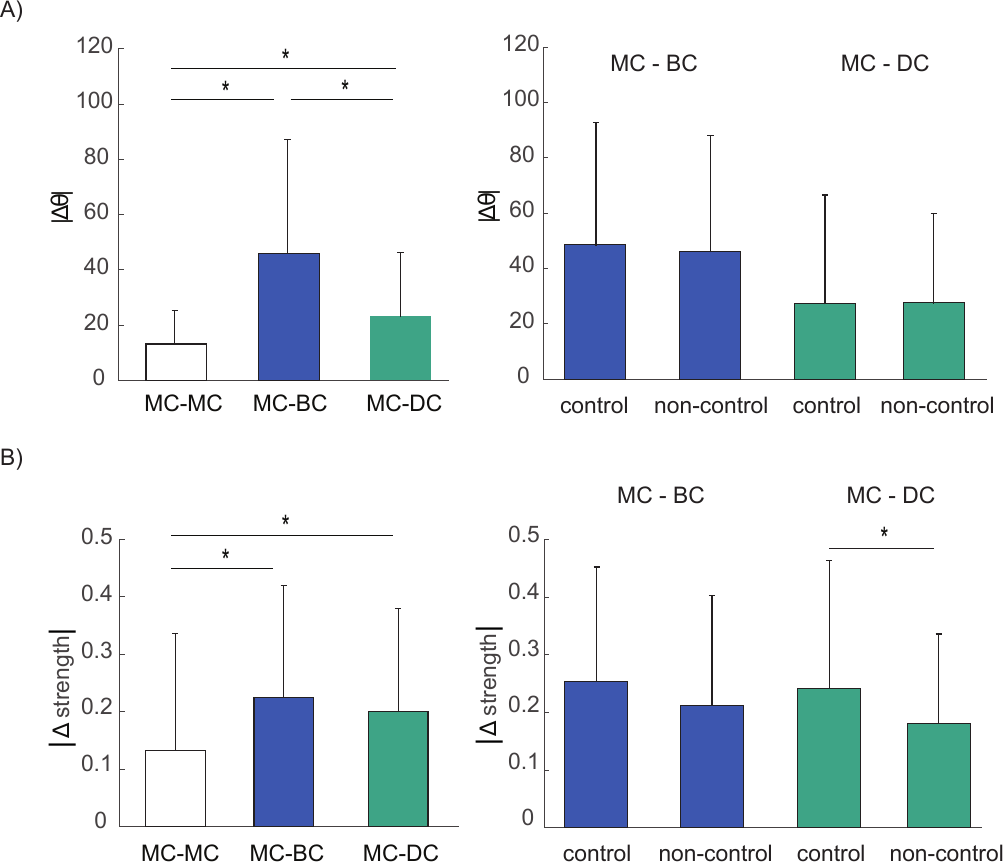}
\caption{Change in tuning angle between conditions for control units that were displayed in their preferred direction on the screen, rather than rotating by 90 degrees as in the remainder of the study. Changes are similar to those observed in the 90 degree rotated dataset (compare with Fig. 2). Overall larger changes are observed between manual control and brain control tasks than manual control and dual control tasks. A) Tuning angle. Left: all $p \ll 0.001$ (two sided t-tests). Right: non-significant. B) Tuning strength. Left: MC-BC $p\ll0.001$, MC-DC $p = 0.0016$ (two sided t-tests) Right: MC-DC $p = 0.0491$ (two sided  t-test). MC = manual control, BC = brain control, DC = dual control, and MC2 = post-manual control session.}
\label{figS3}
\end{figure}

Refer to Fig. \ref{figS3}.

\section*{Acknowledgment}

The authors thank Paul House for expert array implantations, Rob Robinson for animal handling and data collection, Charlie Matlack for task design, and Larry Shupe for technical support. This work was supported by an American Heart \& Stroke Association Scientist Development Grant [NCRP 09SDG2230091]; The Paul G. Allen Family Foundation (Allen Distinguished Investigator Award); National Institutes of Health [NS12542, RR00166]; the Center for Neurotechnology, a National Science Foundation Engineering Research Center [EEC-1028725]; and the WRF UW Institute for Neuroengineering.

\bibliographystyle{IEEEtran}
\bibliography{refs_v2}

\end{document}